\documentclass{aa}
\usepackage{txfonts}
\usepackage{graphicx}

\begin{document}

\title{Water destruction by X-rays in young stellar objects}
\author{P. St\"auber\inst{1} \and J.K. J{\o}rgensen\inst{2} \and E.F. van 
Dishoeck\inst{3} \and S.D. Doty\inst{4} \and A.O. Benz\inst{1}}
\offprints{P. St\"auber, \email{pascalst@astro.phys.ethz.ch}}
\institute{Institute of Astronomy, ETH Z\"urich, CH-8092 Z\"urich, Switzerland 
\and Harvard-Smithsonian Center for Astrophysics, 60 Garden Street, Cambridge, 
MA 02138, USA
\and Sterrewacht Leiden, PO Box 9513, 2300 RA Leiden, The Netherlands 
\and Department of Physics and Astronomy, Denison University, Granville, 
OH 43023, USA}
\date{Received / Accepted}

\abstract{}
{We study the H$_2$O chemistry in star-forming environments under the influence 
of a central X-ray source and a central far ultraviolet (FUV) radiation field. 
The X-ray models are applied to envelopes around low-mass Class $0$ and I 
young stellar objects (YSOs).}
{The gas-phase water chemistry is modeled as a function of time, hydrogen 
density and X-ray flux. To cover a wide range of physical environments, 
densities between $n_{\rm H} = 10^4$--$10^9$\,cm$^{-3}$ and temperatures 
between $T=10$--$1000$\,K are studied.}
{Three different regimes are found: For $T<100$\,K, the water abundance is 
of order $10^{-7}$--$10^{-6}$ and can be somewhat enhanced or reduced due to 
X-rays, depending on time and density. For $100$\,K $\lesssim T \lesssim 
250$\,K, H$_2$O is reduced from initial $x({\rm H_2O}) \approx 10^{-4}$ 
following ice evaporation to $x({\rm H_2O}) \approx 10^{-6}$ for $F_{\rm X} 
\gtrsim 10^{-3}$\,ergs\,s$^{-1}$\,cm$^{-2}$ ($t=10^4$\,yrs) and for $F_{\rm X} 
\gtrsim 10^{-4}$\,ergs\,s$^{-1}$\,cm$^{-2}$ ($t=10^5$\,yrs). At higher 
temperatures ($T \gtrsim 250$\,K) and hydrogen densities, water can persist with $x({\rm 
H_2O}) \approx 10^{-4}$ even for high X-ray fluxes. Water is destroyed in both 
Class $0$ and I envelopes on relatively short timescales ($t \approx 
5000$\,yrs) for realistic X-ray fluxes, although the effect is less prominent 
in Class $0$ envelopes due to the higher X-ray absorbing densities there. FUV 
photons from the central source are not effective in 
destroying water.}
{X-rays reduce the water abundances especially in regions where the gas 
temperature is $T \lesssim 250$--$300$\,K for fluxes $F_{\rm X} \gtrsim 
10^{-5}$--$10^{-4}$\,ergs\,s$^{-1}$\,cm$^{-2}$. The affected regions can be 
envelopes, disks or outflow hot spots. The average water abundance in Class I 
sources for $L_{\rm X} \gtrsim 10^{27}$\,ergs\,s$^{-1}$ is predicted to be 
$x({\rm H_2O}) \lesssim 10^{-6}$. Central  UV fields have a negligible 
influence, unless the photons can escape through cavities.}

\keywords{Stars: formation -- ISM: molecules -- X-rays: ISM -- Molecular 
processes -- Astrochemistry}

\maketitle

\titlerunning{Water destruction by X-rays in young stellar objects}
\authorrunning{P. St\"auber et al.}

\section{Introduction}
\label{pintro}

The importance of water in star-formation is unquestioned. Water is found 
abundantly on dust grains as well as in the gas-phase of molecular clouds, 
cores, envelopes and protostellar disks. It is a key molecule in the hot ($T 
\gtrsim 100$\,K) gas-phase chemistry and is one of the main gas coolants. The 
evolution over time of molecular cores influences the abundance of H$_2$O 
(Ceccarelli et al. \cite{ch96}), making it sensitive to the chemical age of a 
source. Through its masing ability, water has been used to study dynamical 
properties of star-forming regions.

In cold gas ($T<100$\,K), water is mainly produced by ion-molecule 
reactions. At temperatures above $250$\,K, H$_2$O is largely produced 
in reactions of O and OH with H$_2$. Water is also efficiently generated on 
dust grains where it forms an icy mantle. Observed abundances of water 
ice are typically $x({\rm H_2O}) \approx 10^{-4}$ with respect to H$_2$ 
(Tielens et al. \cite{tt91}; Boogert et al. \cite{bp04}). At temperatures 
$T \gtrsim 100$\,K, H$_2$O evaporates into the gas-phase and has a major impact 
on the chemistry (van Dishoeck \& Blake \cite{vb98}; Doty et al. \cite{dv02}). 
Water is also believed to be formed in shocked regions where it may be produced 
due to the high gas temperature and/or may be released from grains. Towards 
high-mass star-forming regions, hot ($T \gtrsim 100$\,K) gas-phase water is 
usually observed with abundances $x({\rm H_2O}) = 10^{-4}$, comparable to 
the ice abundance -- whereas at temperatures below $100$\,K, it is found to be 
approximately $100$ times less abundant (e.g., Boonman \& van Dishoeck 
\cite{bv03}, van der Tak et al. \cite{vw05}). The observed water emission 
towards low-mass objects has been interpreted with similar jumps at $T \approx 
100$\,K (Ceccarelli et al. \cite{cc00}; Maret et al. \cite{mc02}). Although the 
solid water abundances are comparable in low and high-mass regions, the hot 
gas-phase water in low-mass sources seems to be much less abundant than in the 
massive objects. In addition, a fraction of the emission is believed to arise 
in outflow regions, as inferred from observations both on and off source with 
the Long Wavelength Spectrometer (LWS) on board the Infrared Space Observatory 
(ISO) (Giannini et al. \cite{gn01}; Benedettini et al. \cite{bv02}). Nisini et 
al. (\cite{ng02}) compared the far-infrared spectra of both Class $0$ and Class 
I sources. Water lines were found to be prominent in the spectra of Class $0$ 
objects but they were not detected toward Class I sources corresponding 
to an upper limit on the abundances of $x({\rm H_2O}) \lesssim 10^{-5}$.

Protostars are often found to be sources of strong X-ray fluxes (e.g., Casanova 
et al. \cite{cm95}; Koyama et al. \cite{kh96}; Imanishi et al. \cite{ik01}). 
The observed X-rays are understood to be thermal emission from magnetic stellar 
activities or from the disk-star system (e.g., Feigelson \& Montmerle 
\cite{fm99}). Typical X-ray luminosities range from $L_{\rm X} \approx 
10^{28}$--$10^{31}$\,ergs\,s$^{-1}$ in the $0.5$--$10$\,keV band with plasma 
temperatures between $0.6$--$7$\,keV (e.g., Imanishi et al. \cite{ik01}). 
Flares can even lead to plasma temperatures temporarily exceeding $10^8$\,K 
with luminosities higher than $L_{\rm X} \approx 10^{32}$\,ergs\,s$^{-1}$. 
Within the Chandra Orion Ultradeep Project (COUP), Preibisch et al. 
(\cite{pk05}) found a ratio for the X-ray luminosity to the bolometric 
luminosity of ${\rm log}(L_{\rm X}/L_{\rm bol}) \approx -3.6$ for Class I and 
older protostars. No X-ray detection has been reported to date toward Class $0$ 
objects (Hamaguchi et al. \cite{hc05a}). However, this might be due to the 
large X-ray absorbing hydrogen and dust column densities found in these 
sources. The nature of the sources in recent reports of X-ray detections 
towards very young objects is poorly known and they could be either Class $0$ 
or I objects (Hamaguchi et al. \cite{hc05b}; Forbrich et al. \cite{fp05}).

The influence of a central X-ray source on the envelope around young stellar 
objects has recently been studied by St\"auber et al. (\cite{sd05}). It was 
found that the ionization rate in the inner region of the surrounding envelope 
may be dominated by the ionizing X-ray flux rather than by an inner far 
ultraviolet (FUV) radiation field or the cosmic-ray ionization rate. The 
ionizing flux enhances H$_3^+$ and He$^+$ that trigger a distinct chemistry. It 
was also found that the H$_2$O abundances in the gas-phase may be reduced due 
to X-rays. The abundance and behavior of H$_2$O in both the solid and the 
gas-phase will be studied extensively with the upcoming Herschel Space 
Observatory. It is therefore timely to investigate the influence of X-rays and 
inner FUV fields on the abundances of gas-phase water in more detail. 

In this paper, we first study the general influence of X-rays on gas-phase 
water. The time-dependent chemical X-ray models of St\"auber et al. 
(\cite{sd05}) are used and applied to different density and temperature regimes 
for different X-ray fluxes (Sect.~\ref{pgps}). The densities and temperatures 
are chosen to cover a wide range of conditions that are applicable to cold 
molecular clouds, envelopes, protostellar disks and outflow hot spots. Regions 
with density and temperature gradients are modeled and discussed in 
Sect.~\ref{ptmc1}. These regions are representative of low-mass Class $0$ and 
I envelopes. In addition to X-rays, central FUV fields are considered. The aim 
is to find systematic trends between the different type of sources. The results 
for the different envelopes are discussed in Sect.~\ref{pdis}. The conclusions 
are drawn in Sect.~\ref{pc}.


\section{General parameter study}
\label{pgps}

The gas-phase water abundance is modeled as a function of time, hydrogen 
density ($n_{\rm H} = n({\rm H}) + 2n({\rm H_2})$) and X-ray flux. The results 
of this study can be applied to a wide range of different physical environments 
including protostellar envelopes and disks.

\subsection{Chemical model}
\label{pmc}

The X-ray chemistry models are described in detail by St\"auber et al. 
(\cite{sd05}). The models are an extension of the gas-phase chemical models of 
Doty et al. (\cite{dv02}, \cite{ds04}) to allow the impact of X-rays from the 
central source on the surrounding molecular envelope. The input parameters for 
the model are the initial molecular or atomic abundances, the hydrogen density, 
the gas temperature, the X-ray luminosity, the X-ray emitting plasma 
temperature, the X-ray absorbing hydrogen column density, the cosmic-ray 
ionization rate and the enhancement of the inner and outer FUV field with 
respect to the average interstellar radiation field (ISRF). 

The chemical model is based on the UMIST gas-phase chemical reaction network 
(Millar et al. \cite{mf97}) and calculates the time-dependent number density 
$n({\rm i})$ of each species for a certain temperature and distance from the 
source. For the initial chemical abundances we follow the models of Doty et al. 
(\cite{dv02}, \cite{ds04}) for the high-mass source AFGL $2591$ and low-mass 
source IRAS $16293$--$2422$. These models successfully reproduced many of the 
observed molecular lines. The effects of evaporation of a certain species into 
the gas-phase have been approximated by initially depleting this species below 
its evaporation temperature $T_{\rm ev}$. Specifically, all H$_2$O is assumed 
to evaporate at $T > 100$\,K. No photodesorption of ice is taken into account. 
The initial abundances are listed in Table~\ref{tinit}. The adopted cosmic-ray 
ionization rate is discussed in Sect.~\ref{pmp}. 

The gas temperature has been taken to be closely coupled to the dust 
temperature. Although the X-ray flux and the gas temperature are not fully 
independent, they have been treated as uncoupled variables in order to study 
the dependence of H$_2$O on each of them. To cover a wide range of different 
physical environments, the hydrogen density is varied between $n_{\rm H} = 
10^4$--$10^9$\,cm$^{-3}$. At higher densities, the mean free path of X-rays 
becomes small (a few AU). Typical densities in the inner part of envelopes or 
protoplanetary disk atmospheres are between $n_{\rm H} = 
10^6$--$10^9$\,cm$^{-3}$, whereas those for outflows are $n_{\rm H} = 
10^4$--$10^6$\,cm$^{-3}$. The adopted gas temperature is between 
$10$--$1000$\,K. At higher temperatures ($T \gtrsim 2000$\,K), molecular 
hydrogen and water are dissociated due to H and H$_2$ collisions. The modeled 
X-ray fluxes are between $F_{\rm X} = 
10^{-6}$--$10$\,ergs\,s$^{-1}$\,cm$^{-2}$. The initial X-ray spectrum is 
reduced preferentially at low energies by the intervening hydrogen column 
density such that the local X-ray flux roughly scales with 
\begin{equation}
F_{\rm X} \approx 
\biggl(\frac{L_{\rm X}}{10^{31}\,{\rm ergs\,s^{-1}}}\biggr) 
\biggl(\frac{N_{\rm H}}{5.0\times 10^{21}\,{\rm cm^{-2}}}\biggr)^{-\gamma} 
\biggl(\frac{r}{56\,{\rm AU}}\biggr)^{-2} 
\end{equation}
in ergs\,s$^{-1}$\,cm$^{-2}$, where $L_{\rm X}$ is the original X-ray 
luminosity, $N_{\rm H}$ the X-ray absorbing column density and $r$ is the 
distance from the source. The X-ray fluxes are accurate within $20$\% for 
$T_{\rm X} = 3\times 10^7$\,K and $N_{\rm H} = 5\times 10^{21}$--$10^{23}$\,cm$^{-2}$ with $\gamma=0.3$, for 
$N_{\rm H} = 2$--$5\times 10^{23}$\,cm$^{-2}$ with $\gamma=0.4$ and for $N_{\rm 
H} = 6\times 10^{23}$--$10^{24}$\,cm$^{-2}$ with $\gamma=0.5$. A column density 
of $N_{\rm H} = 5\times 10^{21}$\,cm$^{-2}$ absorbs all photons below $\approx 
1$\,keV (St\"auber et al. \cite{sd05}). The X-ray spectrum is assumed to be 
thermal with a plasma temperature $T_{{\rm X}} = 3 \times 10^7$\,K 
($2.6$\,keV). The shape of the X-ray spectrum, however, has only little 
influence on the chemistry (Maloney et al. \cite{mh96}; St\"auber et al. 
\cite{sd05}). FUV fields are neglected.

All reaction rates that are relevant for the H$_2$O chemistry 
(Sect.~\ref{prnet}) are given with a maximum uncertainty of $25$\% by the UMIST 
database\footnote{http://www.rate99.co.uk}. Self-shielding of CO has been 
included by using the shielding functions of Lee et al. (\cite{lh96}).

\begin{table}
\centering
\caption[]{Initial abundances and cosmic-ray ionization rate.}
\label{tinit}
\begin{tabular}{lll} \hline\hline
Species & Initial abundance & Ref. \\ \hline
Initial abundances $T > 100$\,K \\ 
CO         & 2.0E-04  & a \\ 
CO$_2$     & 3.0E-05  & b \\ 
H$_2$O     & 1.5E-04  & c \\
H$_2$S     & 1.0E-08  & d \\
H$_2$CO    & 8.0E-08  & d \\
N$_2$      & 7.0E-05  & e \\
CH$_4$     & 1.0E-07  & e \\
C$_2$H$_4$ & 8.0E-08  & e \\
C$_2$H$_6$ & 1.0E-08  & e \\
CH$_3$OH   & 1.5E-07  & d \\
O          & 0.0      & e \\
S          & 0.0      & e \\

Initial abundances $T < 100$\,K \\ 
CO  & 2.0E-04  & d \\ 
CO$_2$     & 0.0      & f \\ 
H$_2$O     & 0.0      & f \\
H$_2$S     & 0.0      & f \\
N$_2$      & 7.0E-05  & e \\
CH$_4$     & 1.0E-07  & e \\
C$_2$H$_4$ & 8.0E-08  & e \\
C$_2$H$_6$ & 1.0E-08  & e \\
H$_2$CO $(60 < T($K$) < 100)$ & 8.0E-08 & d \\
H$_2$CO $(T($K$) < 60)$ & 0.0 & d \\
CH$_3$OH $(60 < T($K$) < 100)$ & 1.5E-07 & d \\
CH$_3$OH $(T($K$) < 60)$ & 0.0 & d \\
O          & 1.0E-04  & d \\
S          & 6.0E-09  & g \\

Cosmic-ray ionization rate $\zeta_{{\rm cr}}$ (s$^{-1}$) & 0.8E-17 & h \\ 
\hline
\end{tabular}
\begin{list}{}{}
\item[] All abundances are relative to molecular hydrogen. 
$^a$ J{\o}rgensen et al. (\cite{js02}), $^b$ Boonman et al. (\cite{be03}), 
$^c$ Boonman \& van Dishoeck (\cite{bv03}), $^d$ Doty et al. (\cite{ds04}), 
$^e$ Charnley (\cite{ch97}), $^f$ assumed to be frozen-out or absent in cold 
gas-phase, $^g$ Doty et al. (\cite{dv02}), $^h$ see text.
\end{list}
\end{table}

\subsection{Results}

\begin{figure}
\centering
\resizebox{\hsize}{!}{\includegraphics{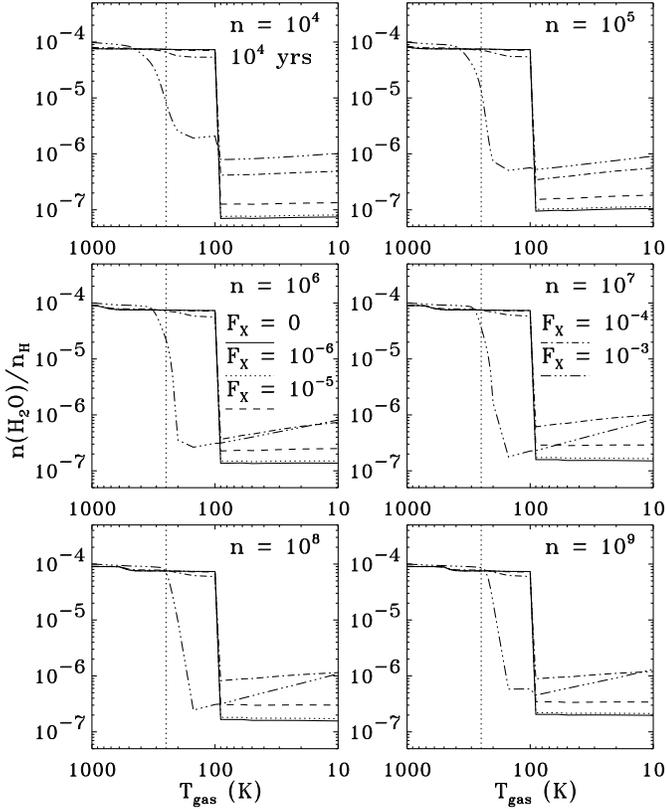}}
\caption{Fractional water abundances as functions of the gas temperature and 
low X-ray flux ($F_{\rm X} = 10^{-6}$--$10^{-3}$\,ergs\,s$^{-1}$\,cm$^{-2}$) 
for different densities (cm$^{-3}$) at $t = 10^4$\,yrs. The vertical line 
indicates the $250$\,K temperature mark.}
\label{ltnlx93}
\end{figure}

\begin{figure}
\centering
\resizebox{\hsize}{!}{\includegraphics{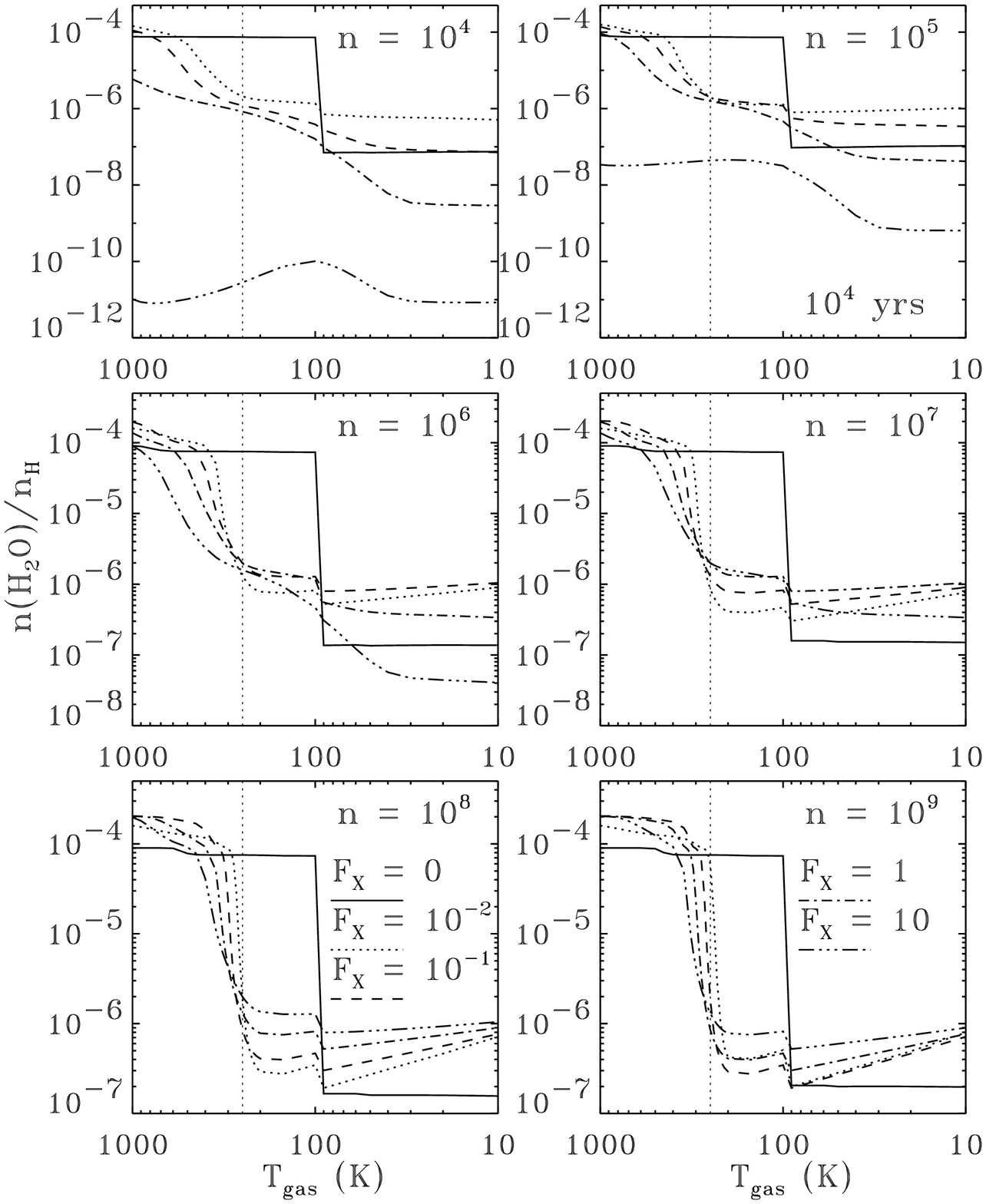}}
\caption{Fractional water abundances as functions of the gas temperature and 
high X-ray flux ($F_{\rm X} = 10^{-2}$--$10$\,ergs\,s$^{-1}$\,cm$^{-2}$) for 
different densities (cm$^{-3}$) at $t = 10^4$\,yrs. The vertical line indicates 
the $250$\,K temperature mark.}
\label{ltnhx93}
\end{figure}

\begin{figure}
\centering
\resizebox{\hsize}{!}{\includegraphics{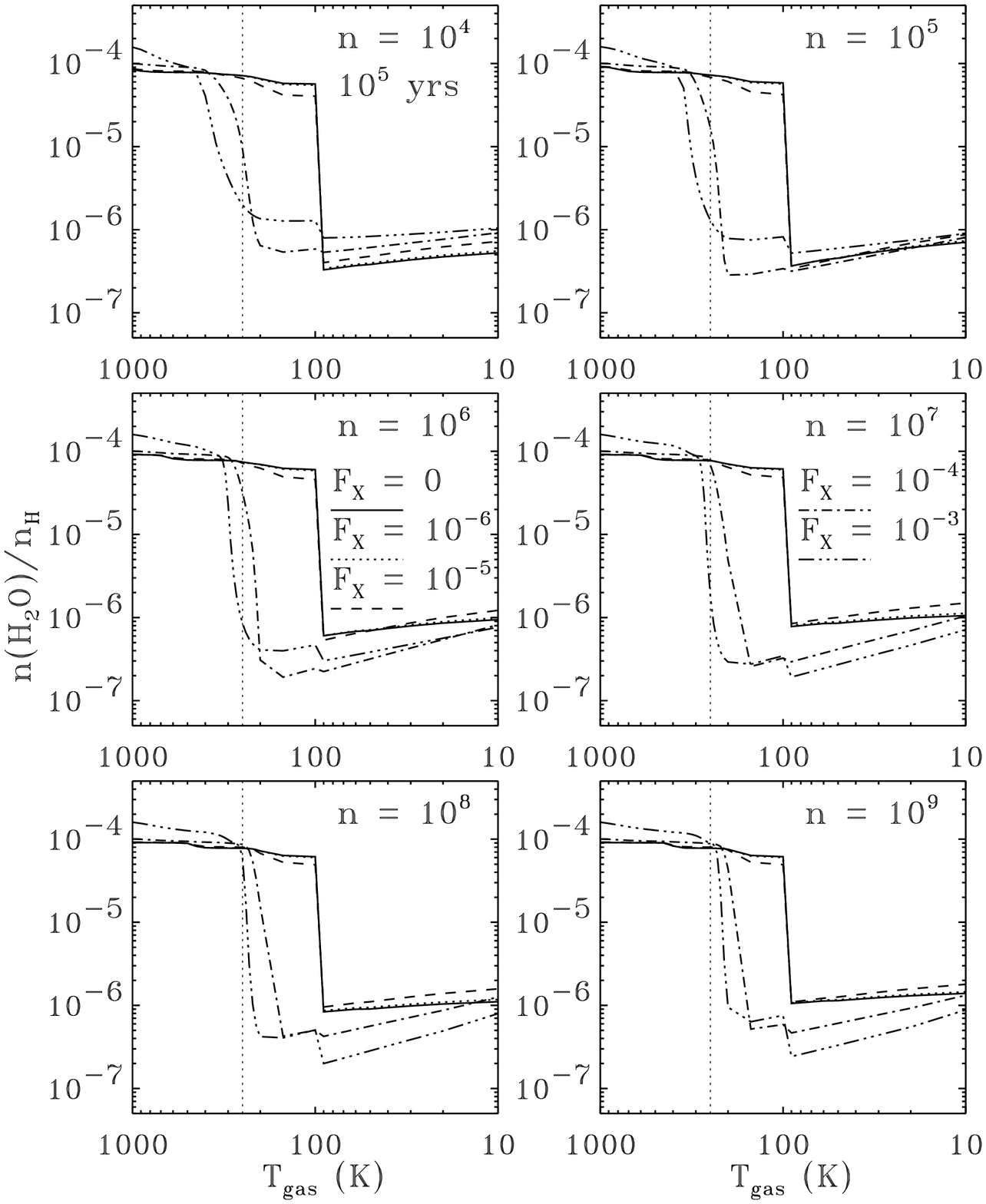}}
\caption{Fractional water abundances as functions of the gas temperature and 
low X-ray flux ($F_{\rm X} = 10^{-6}$--$10^{-3}$\,ergs\,s$^{-1}$\,cm$^{-2}$) 
for different densities (cm$^{-3}$) at $t = 10^5$\,yrs. The vertical line 
indicates the $250$\,K temperature mark.}
\label{ltnlx101}
\end{figure}

\begin{figure}
\centering
\resizebox{\hsize}{!}{\includegraphics{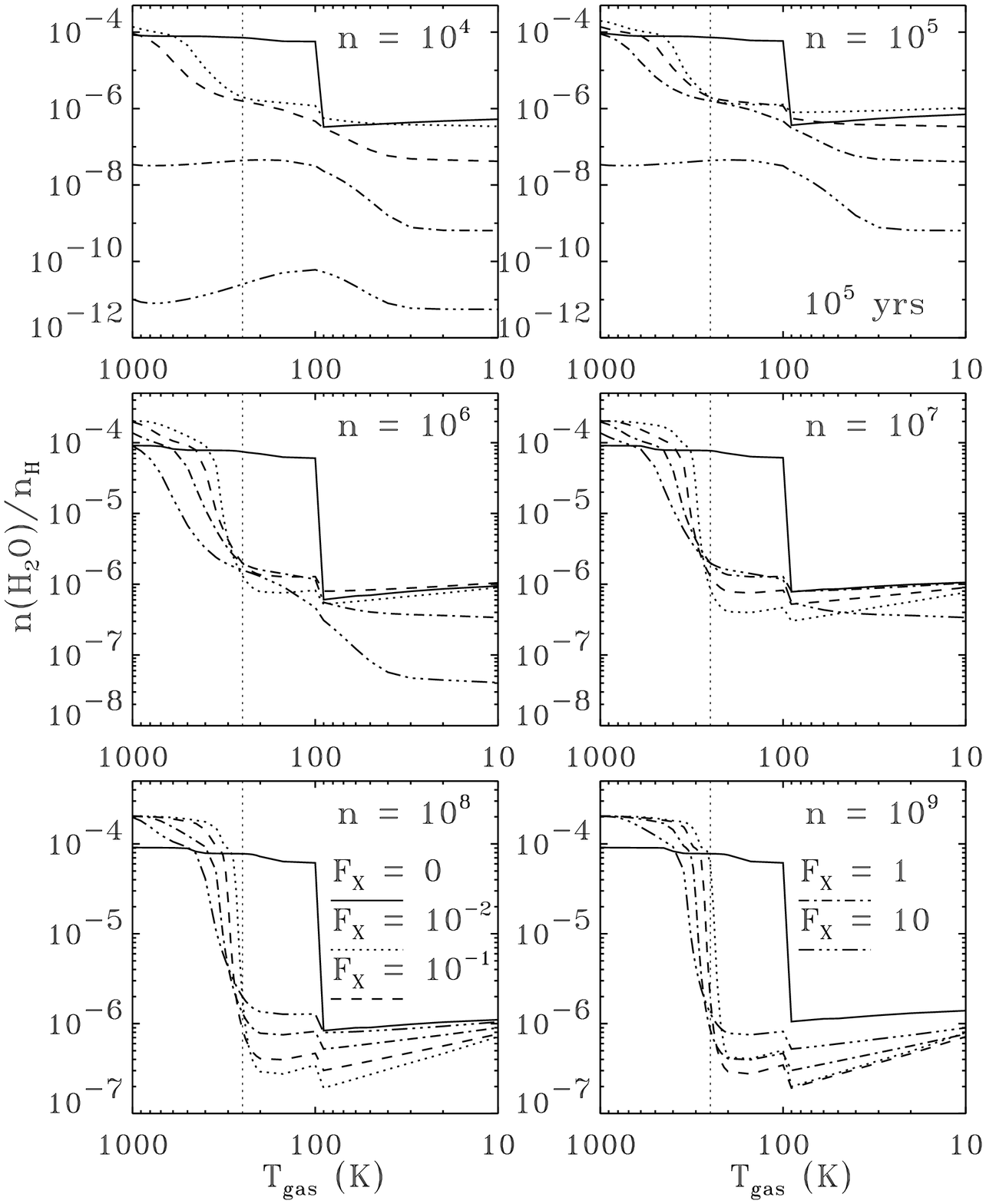}}
\caption{Fractional water abundances as functions of the gas temperature and 
high X-ray flux ($F_{\rm X} = 10^{-2}$--$10$\,ergs\,s$^{-1}$\,cm$^{-2}$) for 
different densities (cm$^{-3}$) at $t = 10^5$\,yrs. The vertical line indicates 
the $250$\,K temperature mark.}
\label{ltnhx101}
\end{figure}

The model results for different X-ray fluxes and hydrogen densities are shown 
in Figs.~\ref{ltnlx93} and \ref{ltnhx93} for $t = 10^4$\,yrs and in 
Figs.~\ref{ltnlx101} and \ref{ltnhx101} for $t = 10^5$\,yrs. It can be seen 
in Figs.~\ref{ltnlx93} and \ref{ltnlx101} that X-ray fluxes $F_{\rm X} 
\lesssim 10^{-6}$\,ergs\,s$^{-1}$\,cm$^{-2}$ have no or minor influence on the 
water abundances for the densities considered. For higher X-ray fluxes, three 
characteristic regimes can be distinguished: $1.$ the regime at $T<100$\,K, 
where gas-phase water is mainly formed and destroyed in ion-molecule reactions 
($x({\rm H_2O}) \approx 10^{-7}$--$10^{-6}$), $2.$ the regime $100$\,K 
$\lesssim T \lesssim 250$\,K, where water is released from grains, but 
destroyed by X-rays ($x({\rm H_2O}) \approx 10^{-6}$) and $3.$ the regime $T 
\gtrsim 250$\,K, where the water abundance is $x({\rm H_2O}) \gtrsim 10^{-4}$. 
The three regimes are discussed in some detail in the following paragraphs.

At $t = 10^4$\,yrs, X-ray fluxes $10^{-5}$ $\lesssim 
F_{\rm X} \lesssim 10^{-2}$\,ergs\,s$^{-1}$\,cm$^{-2}$ enhance the water 
abundance for temperatures $T<100$\,K compared to the model without 
X-rays. The reason for this is the increased H$_3$O$^+$ abundance due to 
X-rays. H$_3$O$^+$ can recombine quickly to H$_2$O in reactions with electrons. 
Note that net formation of H$_2$O at these fluxes can occur because H$_3$O$^+$ 
does not require H$_2$O in its formation. It is produced by ion-molecule 
reactions starting from CO, CO$_2$, O, O$_2$ and OH, which exist in the gas 
phase. Higher X-ray fluxes, however, can enhance water only at densities 
$n_{\rm H} \gtrsim 10^7$\,cm$^{-3}$ for $T<100$\,K (Fig.~\ref{ltnhx93}). At 
$t=10^5$\,yrs, water is only enhanced at $n_{\rm H}=10^4$\,cm$^{-3}$ in this 
temperature regime (Fig.~\ref{ltnlx101}). At higher densities, water is less 
abundant in models with X-rays compared to those without X-rays.

For $100$\,K $\lesssim T \lesssim 250$\,K, the initial abundance of water is 
much higher due to ice evaporation but this high abundance rapidly decreases for 
$F_{\rm X} \gtrsim 10^{-3}$\,ergs\,s$^{-1}$\,cm$^{-2}$ at $t = 10^4$\,yrs and 
for $F_{\rm X} \gtrsim 10^{-4}$\,ergs\,s$^{-1}$\,cm$^{-2}$ at $t = 10^5$\,yrs. 
H$_2$O is mainly destroyed in reactions with H$_3^+$ and HCO$^+$ whose 
abundances are enhanced due to the X-rays. Another destruction mechanism is the 
dissociation of H$_2$O by FUV photons created locally by excited H$_2$. The 
rate for the X-ray induced FUV destruction of water is calculated with the 
method and numbers provided by Maloney et al. (\cite{mh96}). It is an adaption 
of the treatment of Gredel et al. (\cite{gl89}) for the effects of internally 
generated FUV photons by cosmic rays. 

In the temperature regime $T \gtrsim 250$\,K, the reaction OH $+$ H$_2$ 
$\rightarrow$ H$_2$O $+$ H can become more efficient than the water destroying 
reactions and H$_2$O has abundances of the order of $10^{-4}$. At low densities 
($n_{\rm H} = 10^4$--$10^5$\,cm$^{-3}$), high X-ray fluxes lead to the 
destruction of water even for high temperatures. Although the gas temperature 
at these high fluxes may be more than $1000$\,K, water does not reach 
fractional abundances higher than $x({\rm H_2O}) \approx 10^{-7}$, since H$_2$O 
is destroyed in collisions with H and H$_2$ at higher temperatures. The water 
abundance is therefore $x({\rm H_2O}) \lesssim 10^{-10}$ for $F_{\rm X} = 
10$\,ergs\,s$^{-1}$\,cm$^{-2}$ and $n_{\rm H} = 10^{4}$\,cm$^{-3}$ and 
$x({\rm H_2O}) \lesssim 10^{-7}$ for $n_{\rm H} = 10^{5}$\,cm$^{-3}$, 
independent of the gas temperature. At higher densities, however, even high 
X-ray fluxes cannot destroy water in the regime $T \gtrsim 250$\,K. At 
densities $n_{\rm H} \gtrsim 10^7$\,cm$^{-3}$, X-rays even enhance the water 
abundance ($x({\rm H_2O}) > 3\times 10^{-4}$) for temperatures $T \gtrsim 
300$--$600$\,K compared to models without X-rays. This enhancement is 
again mainly through recombination reactions of H$_3$O$^+$. 

At later times ($t=10^6$\,yrs), water is destroyed in the regime $100$\,K 
$\lesssim T \lesssim 250$\,K for $F_{\rm X} \gtrsim 
10^{-5}$\,ergs\,s$^{-1}$\,cm$^{-2}$. The other temperature regimes in the X-ray 
models resemble those at $t=10^5$\,yrs. The main differences with the models 
at $t=10^5$\,yrs, however, are the water abundances of the models without 
X-rays. H$_2$O is gradually destroyed in reactions with HCO$^+$ for 
temperatures $T \gtrsim 100$\,K. In models without X-rays, HCO$^+$ is mainly 
formed by cosmic-ray induced reactions. At $t=10^6$\,yrs, water between $T 
\approx 100$\,K and $T \approx 200$\,K is destroyed down to abundances of a few 
$\times 10^{-7}$. The water abundances for $T<100$\,K on the other hand, are 
slightly higher compared to the models at $t=10^5$\,yrs. This is due to the 
relatively slow ion-molecule reactions that need time to build up water. 

\subsection{Chemical reactions relevant for H$_2$O}
\label{prnet}

\begin{figure}
\centering
\resizebox{\hsize}{!}{\includegraphics[angle=-90,width=8.7cm]{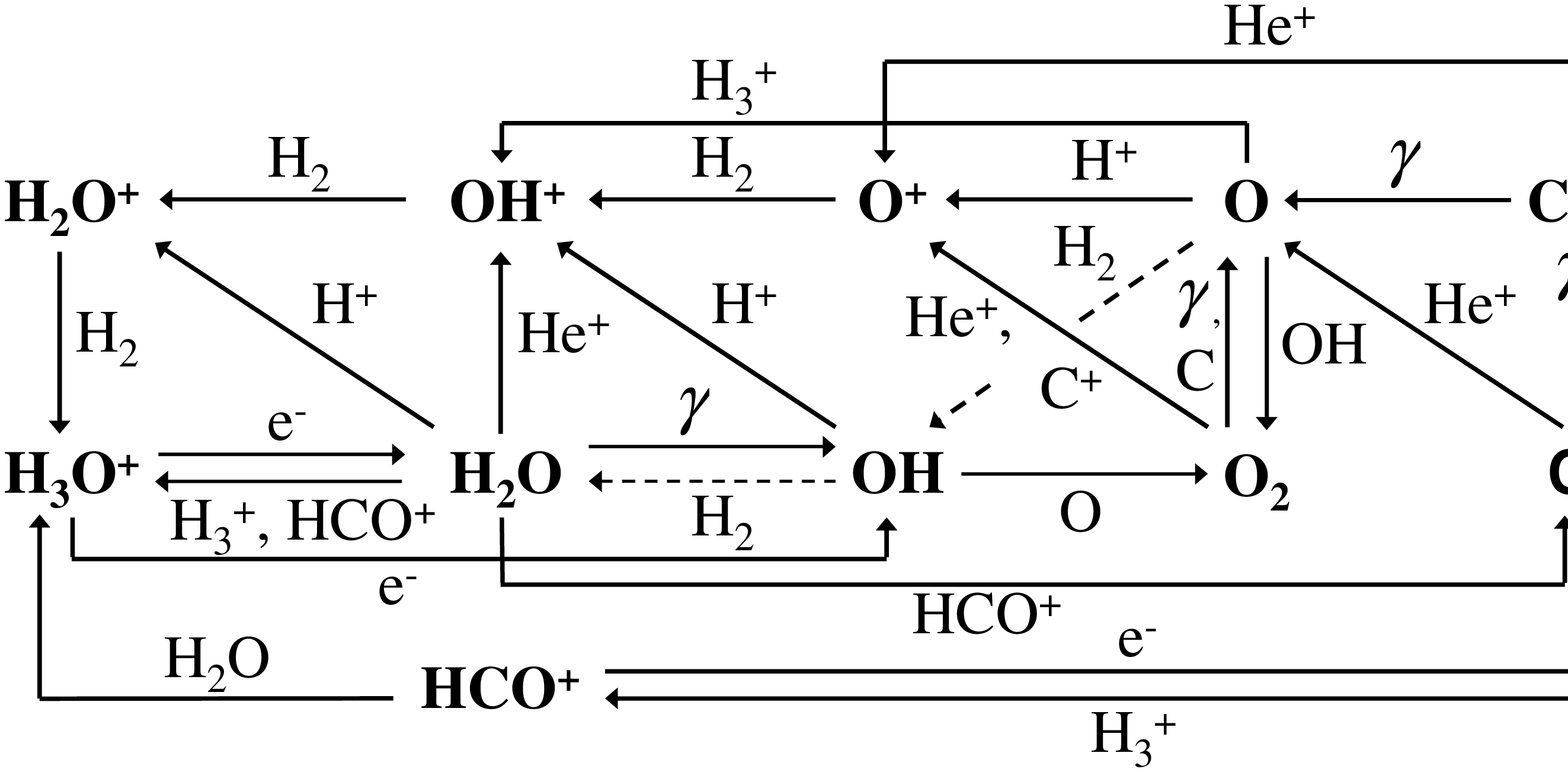}}
\caption{Most important reactions involved in the X-ray chemistry of H$_2$O. 
Dashed lines indicate that the reaction is only important for temperatures $T 
\gtrsim 250$\,K. $\gamma$ denotes the X-ray induced FUV photons. The 
reactions induced by central or interstellar FUV fields are not included.}
\label{fh2onet}
\end{figure}

Water is assumed to be initially frozen out on the dust grains for temperatures 
$T < 100$\,K and injected into the gas at $T = 100$\,K with a fractional 
abundance $x({\rm H_2O}) = 1.5 \times 10^{-4}$. In the presence of X-rays, 
water is then mainly destroyed in reactions with H$_3^+$ and HCO$^+$. Another 
destruction route is the dissociation of water due to X-ray induced FUV 
photons. To estimate the importance of this reaction on the H$_2$O abundance, 
models have been run for different X-ray luminosities ignoring this reaction. 
The resulting H$_2$O abundances vary by less than $15$\% in comparison to models 
where this reaction is included. It emphasizes that the main destruction 
reactions of H$_2$O are the ion-molecule reactions with H$_3^+$ and HCO$^+$, 
respectively.The bulk of the H$_2$O not evaporated from the grains stems from 
the reaction of OH with H$_2$, which becomes dominant for temperatures $T 
\gtrsim 250$\,K, and from electron recombination reactions of H$_3$O$^+$. The 
adopted branching ratios for the dissociative recombination of H$_3$O$^+$ to 
H$_2$O and OH are $0.3$ and $0.7$, respectively (e.g., Jensen et al. 
\cite{jb00}). H$_3$O$^+$ is abundantly produced in the reactions of water with 
H$_3^+$ and HCO$^+$ in the inner region of the envelope. Further away from the 
source, H$_3$O$^+$ is most efficiently formed in the following way: H$_3^+$ 
reacts with atomic oxygen to produce OH$^+$. This is followed by a series of 
hydrogen abstract reactions with H$_2$ leading to H$_3$O$^+$. The main source 
for H$_3^+$ is the reaction of molecular hydrogen with H$_2^+$, which is in 
predominantly produced by electron impact ionization of H$_2$. HCO$^+$ is 
produced via the reaction of H$_3^+$ with CO. OH is largely produced in 
recombination reactions of H$_3$O$^+$ and destroyed in reactions with atomic 
oxygen or molecular hydrogen. The chemical network with reactions and species 
relevant for H$_2$O in the X-ray models is presented in Fig.~\ref{fh2onet}. 
Photoionizations and dissociations due to inner or outer FUV fields are not 
shown.


\section{Applications to protostellar envelopes}								
\label{ptmc1}

In order to study the influence of X-rays on the gas-phase water abundance in 
more realistic environments, the models are applied to envelopes of Class I 
($t \approx 10^5$--$10^6$\,yrs) and Class $0$ ($t \approx$ a few $\times 
10^3$--$10^4$\,yrs) low-mass YSOs. These regions may be characterized by a 
power-law density distribution and an increasing gas temperature towards the 
central protostar. Class $0$ and I envelopes mainly differ in their total 
mass. Typical envelope masses of Class $0$ objects are $M_{\rm env} \gtrsim 
0.5$\,M$_{\sun}$ whereas Class I sources have less massive envelopes with 
$M_{\rm env} \lesssim 0.5$\,M$_{\sun}$ (e.g., J{\o}rgensen et al. \cite{js02}, 
\cite{js05}). Although spherical $1$D models are an approximation to the 
real geometry of protostellar cores, they successfully explain many of the 
observed features of low-mass protostars from few hundred AU out to $10^4$\,AU 
scales (e.g., Sch\"oier et al. \cite{sj04}; J{\o}rgensen et al. \cite{js02}, 
\cite{jb05}) and provide a framework for discussing their overall chemical 
structure (e.g., Doty et al. \cite{ds04}; Maret et al. \cite{mc04}; 
J{\o}rgensen et al. \cite{jj04}). Inside a few hundred AU, however, the 
geometry is likely to be much more complex with cavities and a flattened 
protostellar disk in addition to the spherical inner envelope. Since no 
realistic physical models based on observations on those scales are yet 
available, we adopt the spherical models extrapolated to small radii as a 
starting point. Possible effects due to cavities and non-spherical symmetry are 
discussed throughout this section and in Sect.~\ref{pdis}.

In the following models, the H$_2$O chemistry is assumed to be irradiated 
by X-rays and FUV photons from the central source. The FUV models have been 
described in detail by St\"auber et al. (\cite{sd04}).

\subsection{Class I envelope model}
\label{pmp}

\begin{figure}
\centering
\resizebox{\hsize}{!}{\includegraphics{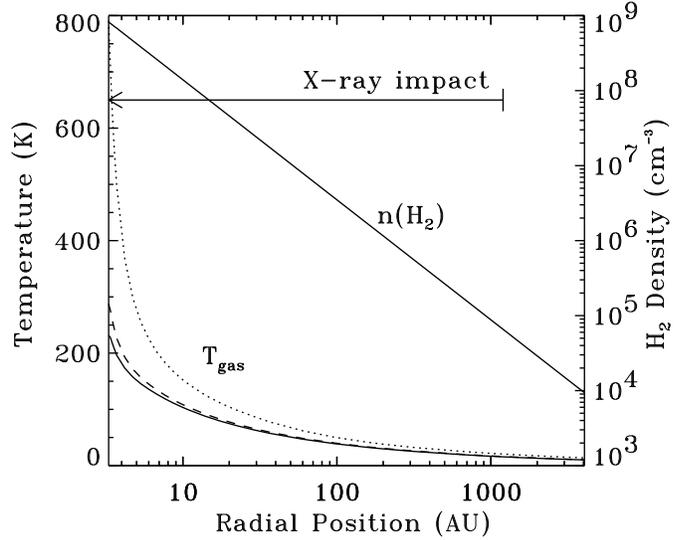}}
\caption{Density and thermal structure for a prototypical Class I object 
adopted from the model results of J{\o}rgensen et al. (\cite{js02}) (solid 
lines). The dashed and dotted lines refer to the gas temperature for the models 
including an additional inner source of X-ray luminosity ($L_{\rm X} = 
10^{31}$\,ergs\,s$^{-1}$ and $L_{\rm X} = 10^{32}$\,ergs\,s$^{-1}$, 
respectively). The arrow indicates the distance out to which X-rays dominate 
the H$_2$ ionization rate over cosmic-rays ($\zeta_{\rm cr} = 0.8 \times 
10^{-17}$\,s$^{-1}$) for $L_{\rm X} = 10^{30}$\,ergs\,s$^{-1}$.}
\label{modelx}
\end{figure}

The density and temperature profiles adopted for a prototypical Class I source 
are presented in Fig.~\ref{modelx} and Table~\ref{tmod}, based on the observations 
by J{\o}rgensen et al. (\cite{js02}). In that analysis, it was assumed that 
the gas temperature equals the dust temperature. X-rays and FUV fields, 
however, are able to increase the gas temperature due to photoelectric heating. 
The gas temperature has therefore been explicitly calculated for the different 
X-ray and FUV fluxes. The X-ray luminosity is varied between $L_{\rm X} = 
10^{26}$--$10^{32}$\,ergs\,s$^{-1}$ and the FUV field from the central source 
($G_{{\rm 0, in}}$) at the inner radius from $10^5$ to $10^8$ with respect to 
the average ISRF. The heating rate due to X-rays is taken from Maloney et al. 
(\cite{mh96}), and for the FUV heating we follow Bakes \& Tielens 
(\cite{bt94}). The heating rates are then compared to the cooling rate due to 
gas-dust collisions provided by Hollenbach \& McKee (\cite{hm89}). At lower 
densities ($n_{\rm H} \approx 10^5$\,cm$^{-3}$) line cooling due to C$^+$, O 
and CO also becomes important (see also Doty \& Neufeld \cite{dn97}), but these 
effects are minor for our purpose since the densities in the regions of 
interest are always $n_{\rm H} \gtrsim 10^6$\,cm$^{-3}$. 

It is found that only X-ray luminosities $L_{\rm X} 
\gtrsim 10^{31}$\,ergs\,s$^{-1}$ lead to a noticeable increase in the gas 
temperature ($T_{\rm gas}-T_{\rm dust} \gtrsim 10$\,K). The results for $L_{\rm 
X} = 10^{31}$\,ergs\,s$^{-1}$ and $L_{\rm X} = 10^{32}$\,ergs\,s$^{-1}$ are 
plotted in Fig.~\ref{modelx}. The temperature of the gas is only affected in 
the inner $\approx 0.1$\,AU by the central FUV field (not shown in 
Fig.~\ref{modelx}). High FUV fields ($G_{{\rm 0, in}} \gtrsim 10^7$) are needed 
though to increase the gas temperature significantly above the dust 
temperature. 

The effects of the outer FUV field can be neglected. The optical depth in the 
region of interest ($T \gtrsim 100$\,K) is $A_{{\rm V}} \approx 18$ and 
all FUV photons should be absorbed by the outer envelope for typical values of 
the ISRF. The outer FUV field is taken to be $G_{{\rm 0, out}} = 1$ according 
to the average ISRF.

The adopted value of the cosmic-ray ionization rate $\zeta_{{\rm cr}} = 0.8 
\times 10^{-17}$\,s$^{-1}$ is based on the results of St\"auber et al. 
(\cite{sd05}) who showed that low-$J$ lines of HCO$^+$ tend to trace the 
cosmic-ray ionization rate, whereas high-$J$ lines are more sensitive to the 
ionizing inner X-ray flux. To constrain $\zeta_{{\rm cr}}$, the rate is varied 
in our models in order to fit the HCO$^+$ $1$--$0$ observations toward the 
low-mass star-forming region Taurus (TMC $1$) of Hogerheijde et al. 
(\cite{hv97}). It is found that $\zeta_{{\rm cr}} = 0.8 \times 
10^{-17}$\,s$^{-1}$ fits the observations best. This value is at the lower end 
of the rates found by van der Tak \& van Dishoeck (\cite{vv00}) toward 
high-mass star-forming regions ($\zeta_{{\rm cr}} = 0.61$--$5.6 \times 
10^{-17}$\,s$^{-1}$).

\begin{table}[]
\centering
\caption[]{Model parameters for a prototypical Class I object.}
\begin{tabular}{ll} \hline\hline
$L_{{\rm bol}}$ & 0.66\,L$_{\sun}$ \\ 
Envelope parameters: &  \\ \hline
Inner radius $r_{\rm in}$ ($T=250$\,K) & 3.1\,AU \\
Outer radius $r_{\rm out}$ ($T=10$\,K) & 4.3 $\times 10^3$\,AU \\
Density at 1000\,AU, $n({\rm H_2})$ & 8.8 $\times 10^4$\,cm$^{-3}$ \\
Density power-law index, $p$ &  1.6 \\
Gas mass $M_{{\rm env}}$ ($T=10$\,K) & 0.034\,M$_{\sun}$ \\ 
Column density $N_{\rm H_2}$ & $6.9\times 10^{22}$\,cm$^{-2}$ \\ \hline
\label{tmod}
\end{tabular}
\end{table}

\begin{figure*}
\centering
\resizebox{\hsize}{!}{\includegraphics{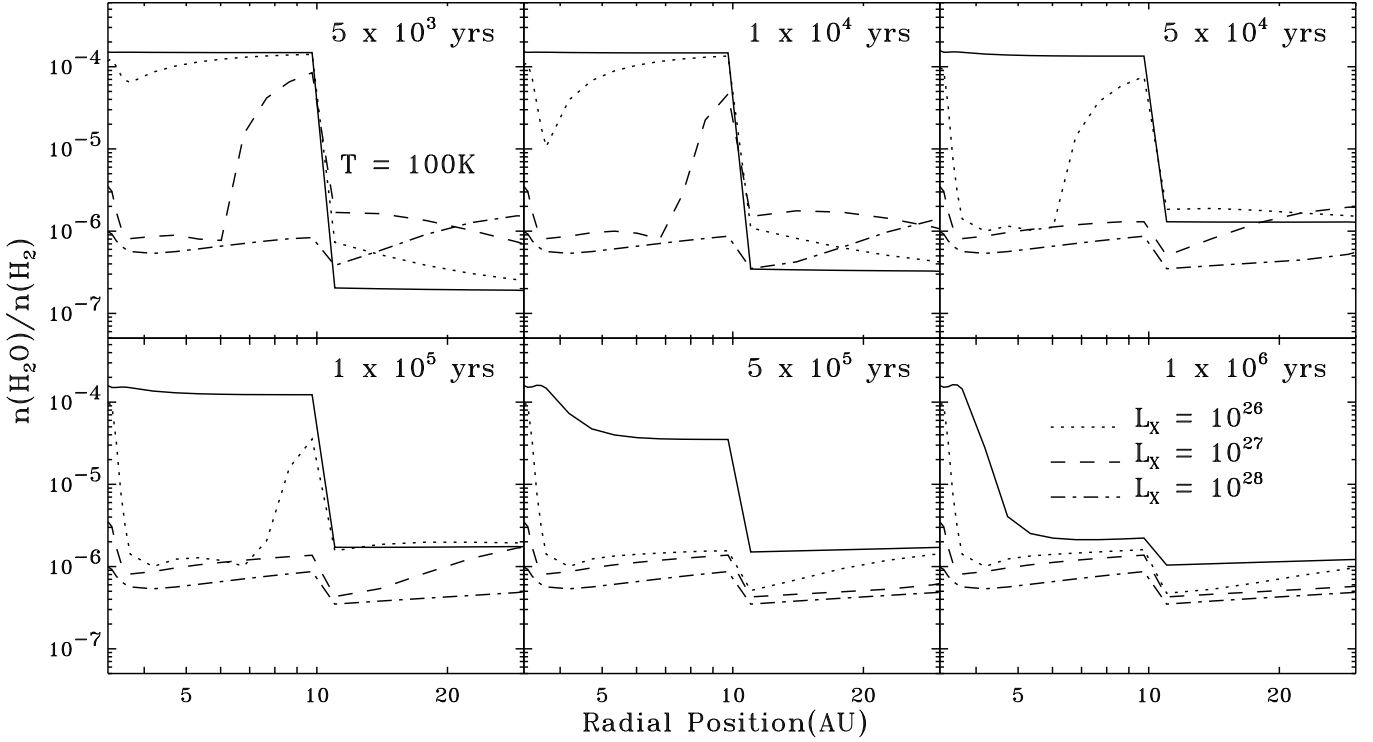}}
\caption{Depth dependent temporal evolution of H$_2$O for different X-ray 
luminosities. The solid line corresponds to the model without X-rays. X-ray 
luminosities are given in ergs\,s$^{-1}$. The $T = 100$\,K mark is at $\approx 
10$\,AU.}
\label{h2ot}
\end{figure*}

\subsection{Results for the Class I model}
\label{pres}

\subsubsection{X-ray models}
\label{prx}

The depth dependent water profiles for $t = 5 \times 10^3$\,yrs to $t = 
10^6$\,yrs and for X-ray luminosities $L_{\rm X} = 
10^{26}$--$10^{28}$\,ergs\,s$^{-1}$ are presented in Fig.~\ref{h2ot}. The 
inner FUV field is neglected in these models. It can be seen that an X-ray 
luminosity of $L_{\rm X} \gtrsim 10^{28}$\,ergs\,s$^{-1}$ destroys the 
gas-phase H$_2$O within less than $5000$\,yrs from initial $x({\rm H_2O}) 
\approx 10^{-4}$ down to $x({\rm H_2O}) \approx 10^{-6}$ with respect to 
molecular hydrogen. For an X-ray luminosity of $L_{\rm X} \approx 
10^{27}$\,ergs\,s$^{-1}$ the timescale to destroy water is $t \gtrsim 5 
\times 10^4$\,yrs. In models with $L_{\rm X} \lesssim 10^{26}$\,ergs\,s$^{-1}$, 
water has still high abundances ($x({\rm H_2O}) \approx 10^{-4}$) in the 
innermost part of the envelope. We therefore conclude that X-ray luminosities 
$L_{\rm X} \gtrsim 10^{27}$\,ergs\,s$^{-1}$ destroy all gas-phase water down to 
fractional abundances of $x({\rm H_2O}) \approx 10^{-6}$ on timescales $t 
\approx 5 \times 10^4$\,yrs and $L_{\rm X} \gtrsim 10^{28}$\,ergs\,s$^{-1}$ on 
a timescale of a few $\times 10^3$\,yrs. 

\begin{figure}
\centering
\resizebox{\hsize}{!}{\includegraphics{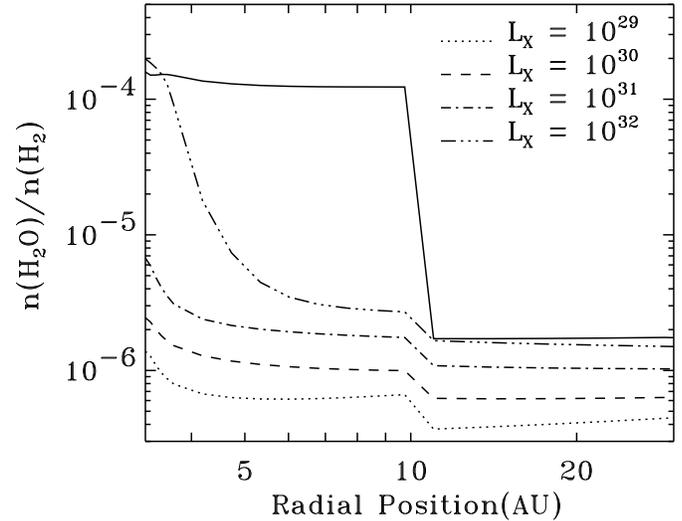}}
\caption{Depth dependent fractional abundances of H$_2$O for high X-ray 
luminosities and $t = 10^5$\,yrs. The solid line corresponds to the model 
without X-rays. X-ray luminosities are given in ergs\,s$^{-1}$.}
\label{h2ohx}
\end{figure}

The water profile for $t = 10^5$\,yrs and higher X-ray luminosities is shown in 
Fig.~\ref{h2ohx}. The fractional abundances increase from $x({\rm H_2O}) 
\approx 10^{-6}$ to $x({\rm H_2O}) \approx 10^{-5}$ for increasing X-ray 
luminosities between $L_{\rm X} = 10^{29}$--$10^{31}$\,ergs\,s$^{-1}$, and 
$x({\rm H_2O})$ reaches even $\gtrsim 10^{-4}$ in the innermost part of the 
envelope for $L_{\rm X} = 10^{32}$\,ergs\,s$^{-1}$. This effect can be 
explained by the fast electron recombination reaction of H$_3$O$^+$ that forms 
H$_2$O and by the increasing gas temperature with increasing X-ray flux: The 
reaction of OH with H$_2$, that drives oxygen into H$_2$O, becomes faster than 
the destruction of water by X-rays at higher temperatures, hence more water is 
produced for higher X-ray fluxes. In the region $T < 100$\,K, $x({\rm H_2O})$ 
decreases with increasing X-ray luminosity for $t = 10^5$\,yrs, until 
$L_{\rm X} \approx 10^{29}$\,ergs\,s$^{-1}$. Higher X-ray luminosities lead to 
higher water abundances. This is again mainly due to the recombination of 
H$_3$O$^+$, which is more abundant in models with high X-ray fluxes (see also 
Sect.~\ref{pgps}). Fig.~\ref{h2ohx} shows that $x({\rm H_2O})$ increases from 
$\approx 4$--$5 \times 10^{-7}$ to $\approx 2 \times 10^{-6}$ with increasing 
X-ray luminosity ($L_{\rm X} \gtrsim 10^{29}$\,ergs\,s$^{-1}$). At early times, 
the recombination of H$_3$O$^+$ in the models with X-rays leads to even higher 
H$_2$O abundances in the region $T < 100$\,K compared with the model without 
X-rays (Fig.~\ref{h2ot}). 

\begin{figure}
\centering
\resizebox{\hsize}{!}{\includegraphics{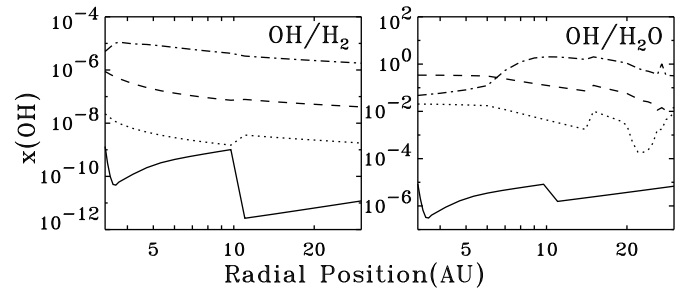}}
\caption{Fractional abundances of OH (left: $x({\rm OH}) = 
n({\rm OH})/n({\rm H_2})$, right: $x({\rm OH}) = n({\rm OH})/n({\rm H_2O})$) 
for the model without X-rays (solid line), for $L_{\rm X} = 
10^{28}$\,ergs\,s$^{-1}$ (dotted line), for $L_{\rm X} = 
10^{30}$\,ergs\,s$^{-1}$ (dashed line) and for $L_{\rm X} = 
10^{32}$\,ergs\,s$^{-1}$ (dashed-dotted line).}
\label{oh2o}
\end{figure}

The OH abundance profiles for different X-ray luminosities at $t = 10^5$\,yrs 
are given in Fig.~\ref{oh2o}. The OH densities are normalized by $n({\rm H_2})$ 
(left) and by $n({\rm H_2O})$ (right), respectively. In the right part of 
Fig.~\ref{oh2o} it can be seen that an X-ray luminosity $L_{\rm X} = 
10^{28}$\,ergs\,s$^{-1}$ leads to $n({\rm OH})/n({\rm H_2O}) \approx 10^{-2}$, 
whereas X-ray luminosities $L_{\rm X} \gtrsim 10^{30}$\,ergs\,s$^{-1}$ can 
produce ratios of $n({\rm OH})/n({\rm H_2O}) \approx 0.1$--$1$ in the innermost 
part of the envelope. The $n({\rm OH})/n({\rm H_2O})$ ratio in the hot ($T 
\gtrsim 100$\,K) region of the envelope in the models without X-rays is only of 
the order of $\approx 10^{-4}$. X-rays therefore enhance the OH abundances as 
well as the OH to H$_2$O abundance ratio.

\begin{figure}
\centering
\resizebox{\hsize}{!}{\includegraphics{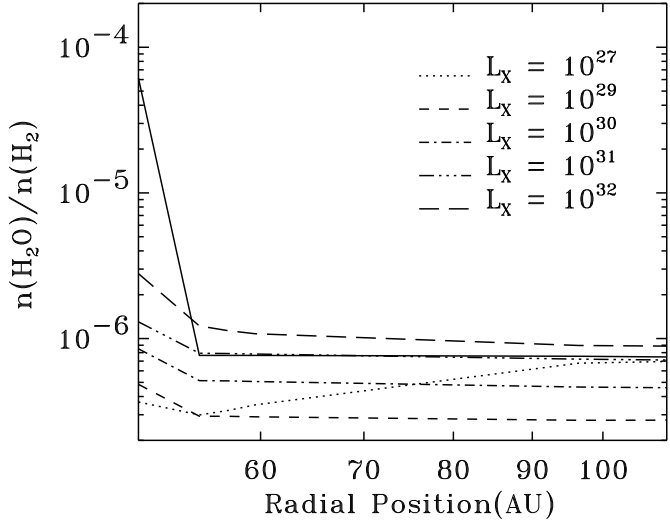}}
\caption{Model with $r_{\rm in} = 50$\,AU and $T(r_{\rm in}) = 100$\,K: Depth 
dependent fractional abundances of H$_2$O for different X-ray luminosities and 
$t = 10^5$\,yrs. The solid line corresponds to the model without X-rays. X-ray 
luminosities are given in ergs\,s$^{-1}$.}
\label{h2o50au}
\end{figure}

The high densities ($n_{\rm H} > 10^8$\,cm$^{-3}$) and temperatures ($T > 
200$\,K) in the model at $r<10$\,AU may be appropriate either to an inner 
envelope or to the inner regions of a circumstellar disk. However, there is 
also evidence for cavities in the inner few hundred AU (e.g., J{\o}rgensen et 
al. \cite{jl05}). We have therefore also modeled an envelope where $r_{\rm in}$ 
is assumed to be at $50$\,AU with a temperature $T = 100$\,K. This corresponds 
to a source with $L_{{\rm bol}} = 2$\,L$_{\sun}$. The results for $t = 
10^5$\,yrs are presented in Fig.~\ref{h2o50au}. H$_2$O is destroyed by the 
X-rays and has abundances $x({\rm H_2O}) < 10^{-5}$. At larger radii 
($r_{\rm in} > 50$\,AU) the dust will not be heated above $T>100$\,K, i.e., the 
temperature where water evaporates into the gas phase, and the water abundances 
will stay below $x({\rm H_2O}) < 10^{-5}$. 

\subsubsection{FUV models}
\label{prfuv}

\begin{figure}
\centering
\resizebox{\hsize}{!}{\includegraphics{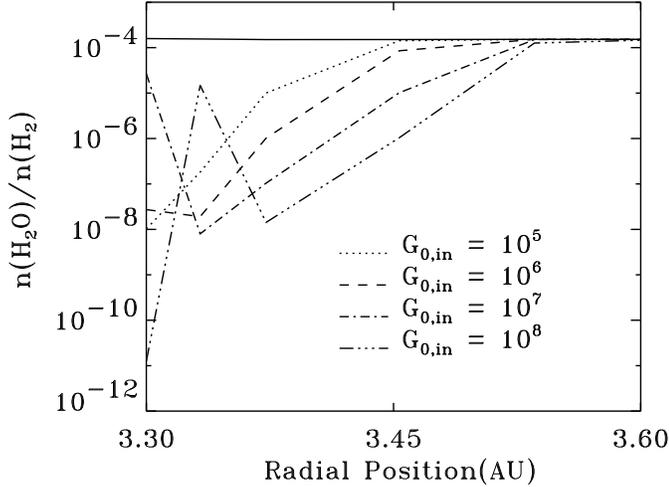}}
\caption{Depth dependent fractional abundances of H$_2$O for different inner 
FUV fields and $t = 10^5$\,yrs. The solid line corresponds to the model with  
$G_{{\rm 0, in}} = 0$. The X-ray emission is neglected.}
\label{h2og}
\end{figure}

The H$_2$O profiles for different inner FUV fields ($G_{{\rm 0, in}} = 
10^5$--$10^8$) are presented in Fig.~\ref{h2og}. The X-ray emission has been 
neglected in these models. As can be seen, inner FUV fields affect the water 
abundances only in the inner $0.2$--$0.3$\,AU. This is due to the high hydrogen 
densities ($n_{{\rm H}} \approx 10^9$\,cm$^{-3}$) in this region that absorb 
the FUV photons very quickly. Although the FUV photons from the central source 
can destroy H$_2$O in the innermost region, even very high FUV fields ($G_{{\rm 
0, in}} = 10^8$) cannot destroy H$_2$O out to $r_{\rm 100\,K} = 10$\,AU. 

The FUV photons may reach distances in the envelope further away from the 
source by traveling through the outflow cones until they are scattered back 
into the envelope by dust grains (Spaans et al. \cite{sh95}; 
J{\o}rgensen \cite{jj04}). However, the influence of the scattered photons is 
restricted to a small region around the outflow walls, since the FUV photons 
are absorbed quickly at these high densities and therefore water will not be 
destroyed on large scales. 

Another possibility for the inner FUV field to affect larger distances might be 
a somewhat clumpy envelope. Models of clumpy photodissociation regions have 
shown that the FUV photons can travel deeper into the molecular cloud (Meixner 
\& Tielens \cite{mt93}). At large radii with lower densities, the FUV photons 
primarily heat the gas and not the dust. H$_2$O desorption will therefore not 
be much affected - but H$_2$O gas may be photodissociated. Spaans \& van 
Dishoeck (\cite{sv01}) have shown that an inhomogeneous density distribution 
in molecular clouds lowers the column density of H$_2$O by more than an order 
of magnitude compared with the homogeneous cloud model. The advent of ALMA or 
space-based far-infrared interferometers might help to investigate the 
possibility of clumpy envelopes in more detail, since a spatial resolution of 
$\lesssim 1$\arcsec\, is needed to resolve the region of interest. 

Models where a central X-ray source and circumstellar FUV fields were combined 
showed similar results for the influence of the FUV photons. It is thus 
concluded that X-rays clearly dominate the chemistry of the envelope in 
comparison to FUV fields.

\begin{figure*}
\centering
\resizebox{\hsize}{!}{\includegraphics{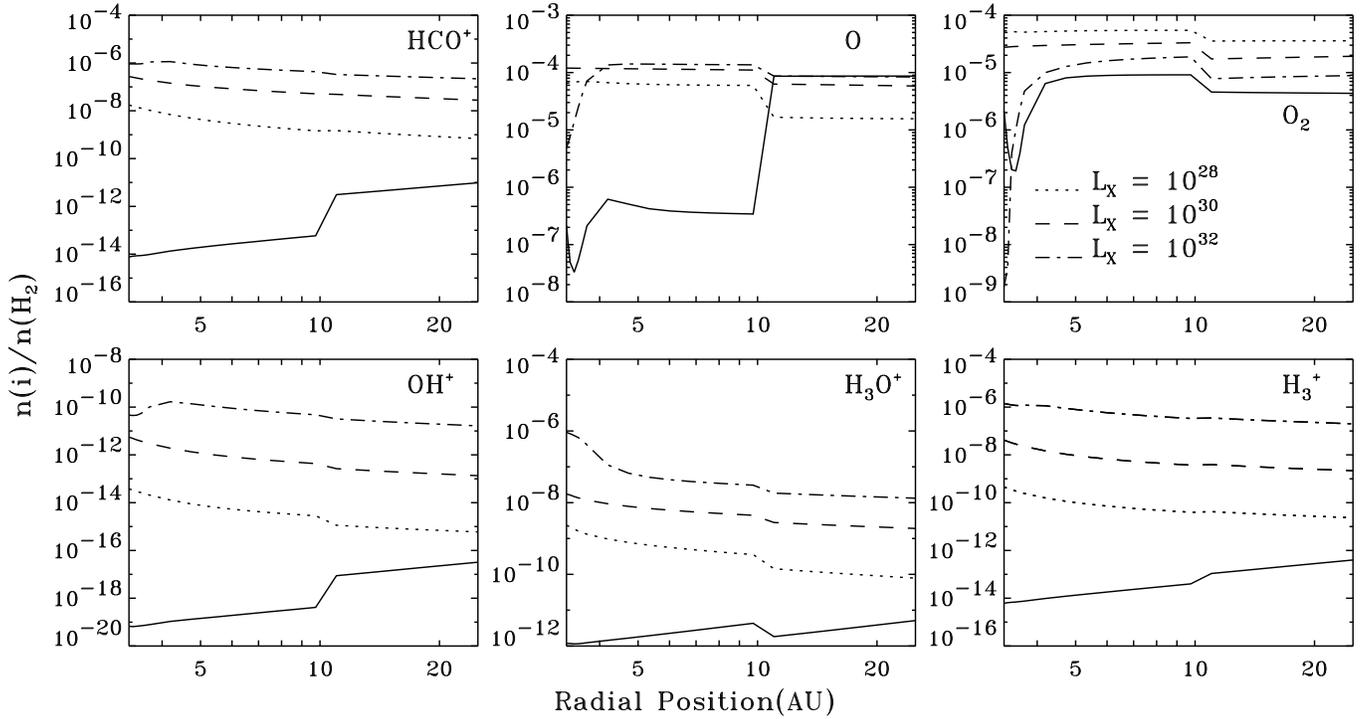}}
\caption{Depth dependent fractional abundances of relevant species in the 
chemical H$_2$O reaction network for different X-ray luminosities and 
$t = 10^5$\,yrs. The solid line corresponds to the model without X-rays or 
inner FUV fields. X-ray luminosities are given in ergs\,s$^{-1}$.}
\label{oplot}
\end{figure*}

\subsubsection{Other species}

Depth dependent fractional abundances of the main species involved in the water 
reaction network are presented in Fig.~\ref{oplot} for $t = 10^5$\,yrs. Atomic 
oxygen is mainly produced by the X-ray induced FUV dissociation of O$_2$ and 
the dissociative ionization of CO by He$^+$. Atomic oxygen reaches fractional 
abundances of $x({\rm O}) \approx 10^{-4}$ in the inner warm ($T > 100$\,K) 
region. The OH$^+$ abundance is dramatically increased by X-rays (see also 
St\"auber et al. \cite{sd05}). The fractional abundance of OH$^+$ is $\approx 
10^{-14}$ for $L_{\rm X} = 10^{28}$\,ergs\,s$^{-1}$, $x({\rm OH^+}) 
\approx 10^{-12}$ for $L_{\rm X} = 10^{30}$\,ergs\,s$^{-1}$ and $x({\rm OH^+}) 
\approx 10^{-10}$ for $L_{\rm X} = 10^{32}$\,ergs\,s$^{-1}$ in the inner region 
of the envelope. It is mainly produced in the ion-molecule reactions of O with 
H$_3^+$, O$^+$ with H$_2$ and OH with H$^+$. 

O$_2$ has fractional abundances $x({\rm O_2}) \approx 10^{-5}$--$10^{-4}$ in 
the models, much higher than observed. Molecular oxygen is efficiently 
produced in the reaction of atomic oxygen with OH. There is an uncertainty in 
the rate of this reaction for O in the ground fine-structure level ($^3P_2$), 
however, and the rate coefficient may be lower than that cited by UMIST (Sims 
et al. \cite{se06}). O$_2$ is mainly destroyed in the reaction with H$_3^+$ 
and by X-ray induced FUV photons. O$_2$ has been searched with SWAS (e.g., 
Melnick \cite{mg04}) and ODIN (e.g., Pagani et al. \cite{po03}) but has not 
been reported in the literature to date. Derived upper limits on O$_2$ column 
densities are usually of the order of $\approx$ a few $\times 
10^{15}$\,cm$^{-2}$ or $x({\rm O_2}) \approx$ a few $\times 10^{-7}$. Standard 
chemical models normally overestimate this value by $2$--$3$ orders of 
magnitude (see Melnick \cite{mg04}, Pagani et al. \cite{po03} for a more 
detailed discussion). Our models are no exception. However, the $30$\arcsec\, 
averaged column density in the Class I models are $N_{\rm beam}({\rm O_2}) 
\lesssim 10^{15}$\,cm$^{-2}$, which is well within the upper limits derived by 
SWAS and ODIN. Oxygen could be frozen out on grains in some form at lower 
temperatures and therefore not seen with the big beams of ODIN ($9$\arcmin) and 
SWAS ($3.5$\arcmin $\times$ $5.0$\arcmin). Herschel might shed some light on 
this matter with its relatively small beam widths ($8$--$35$\arcsec, depending 
on frequency).  

H$_3^+$ is a key molecule in the chemical reaction network of water and its 
abundance is greatly enhanced in the models including X-rays. Thus, 
observations of H$_3^+$ in absorption at $\approx 3.7$\,$\mu$m (e.g., McCall et 
al. \cite{mg99}) would allow a direct test of the models. The integrated radial 
column densities $N_{\rm rad}$ of H$_3^+$ are therefore presented for the 
Class I X-ray models in Table~\ref{th3+}. The values given in parenthesis 
are H$_3^+$ column densities calculated for the model with $r_{\rm in} = 
50$\,AU, i.e. the model with a protostellar hole.

\begin{table}[]
\centering
\caption[]{Radial column densities of H$_3^+$ for the Class I source.}
\begin{tabular}{cccc} \hline\hline
log($L_{\rm X}$) & $N_{\rm rad}$ & log($L_{\rm X}$) & $N_{\rm rad}$ \\
log(ergs\,s$^{-1}$) & cm$^{-2}$ & log(ergs\,s$^{-1}$) & cm$^{-2}$ \\ \hline
0  & 8.0E11 (2.0E09) & 30 & 4.8E14 (4.2E11) \\
27 & 1.3E12 (2.5E09) & 31 & 4.5E15 (3.9E12) \\
28 & 5.9E12 (6.5E09) & 32 & 3.6E16 (2.9E13) \\
29 & 5.0E13 (4.5E10) &    &        \\ \hline
\label{th3+}
\end{tabular}
\end{table}

\subsection{Class $0$ envelope model}
\label{pcl0}

Table~\ref{tiras} lists the main parameters for a prototypical Class $0$ 
envelope model (Sch\"oier et al. \cite{sj02}). The crucial differences from the 
modeling point of view between the Class $0$ and I envelope are the much higher 
bolometric luminosity, envelope mass and column density of the younger object, 
that lead to a different density and temperature distribution. The initial 
chemical abundances are again taken as given in Table~\ref{tinit}. 
Freeze-out of CO is taken into account for temperatures $T<20$\,K. The 
increased gas temperature due to X-rays has been calculated as described in 
Sect.~\ref{pmp}. The X-ray luminosity is varied between $L_{\rm X} = 
10^{28}$--$10^{32}$\,ergs\,s$^{-1}$. 

\begin{figure}
\centering
\resizebox{\hsize}{!}{\includegraphics{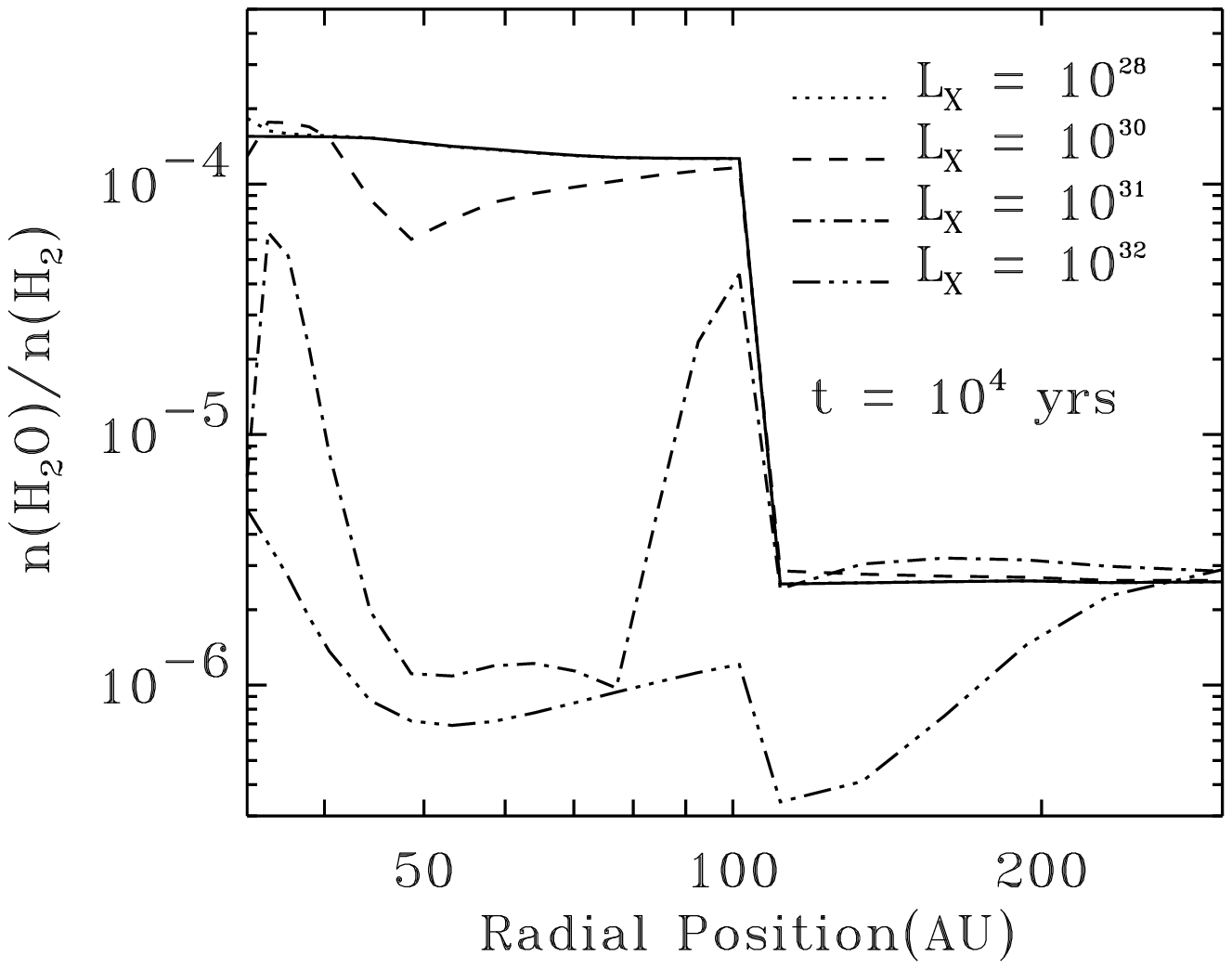}}
\caption{Depth dependent fractional abundances of H$_2$O for a Class $0$ object 
for different X-ray luminosities and $t = 10^4$\,yrs. The solid line corresponds 
to the model without X-rays. X-ray luminosities are given in ergs\,s$^{-1}$.} 
\label{ih2ohx}
\end{figure}

\begin{figure}
\centering
\resizebox{\hsize}{!}{\includegraphics{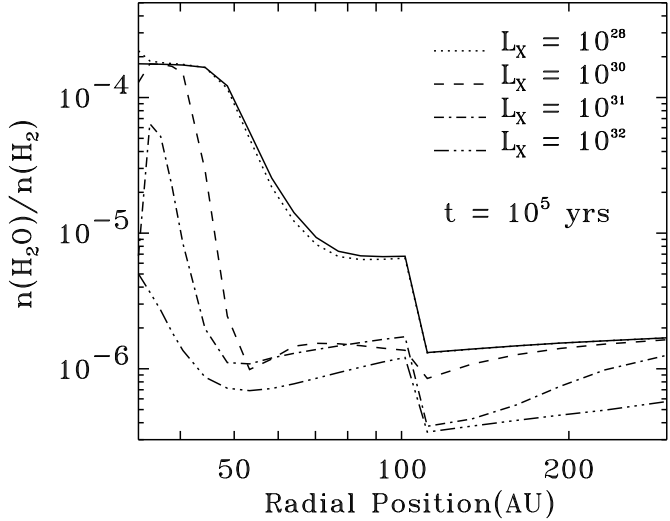}}
\caption{Depth dependent fractional abundances of H$_2$O for a Class $0$ object 
for different X-ray luminosities and $t = 10^5$\,yrs. The solid line corresponds 
to the model without X-rays. X-ray luminosities are given in ergs\,s$^{-1}$.} 
\label{ih2ohx2}
\end{figure}

The results for H$_2$O are shown in Fig.~\ref{ih2ohx} for $t = 10^4$\,yrs and 
in Fig.~\ref{ih2ohx2} for $t = 10^5$\,yrs. At $t = 10^4$\,yrs, only high X-ray 
luminosities ($L_{\rm X} \gtrsim 10^{32}$\,ergs\,s$^{-1}$) lead to water 
abundances $x({\rm H_2O}) \lesssim 10^{-5}$. An X-ray luminosity of $L_{\rm X} 
\gtrsim 10^{31}$\,ergs\,s$^{-1}$ destroys water between $\approx 40$--$80$\,AU 
with higher abundances ($x({\rm H_2O}) \approx$ a few $\times 10^{-5}$) at 
$\approx 35$\,AU and $\approx 100$\,AU. Water is not significantly destroyed 
for X-ray luminosities $L_{\rm X} \lesssim 10^{30}$\,ergs\,s$^{-1}$. Similar 
conclusions hold for $t = 10^5$\,yrs although the region where 
$x({\rm H_2O}) \gtrsim 10^{-5}$ is less than $50$\,AU ($T \gtrsim 250$\,K). 
Compared to the Class I models (after appropriate scaling of the X-ray flux 
$\propto L_{\rm X}r^{-2}$), water is less destroyed by X-rays in the Class $0$ 
envelope. The reason for this are the high densities ($n_{\rm H} \gtrsim 
10^9$\,cm$^{-3}$) in the innermost region of the Class $0$ source that absorb 
the X-ray flux. 

\begin{table}[]
\centering
\caption[]{Model parameters for a prototypical Class $0$ source.}
\begin{tabular}{ll} \hline\hline
$L_{{\rm bol}}$ & 27\,L$_{\sun}$ \\ 
Envelope parameters: &  \\ \hline
Inner radius $r_{\rm in}$ & 32\,AU \\
Outer radius $r_{\rm out}$ & 8 $\times 10^3$\,AU \\
Density at 1000\,AU, $n({\rm H_2})$ & 6.7 $\times 10^6$\,cm$^{-3}$ \\
Density power-law index, $p$ & 1.7 \\
Mass $M_{{\rm env}}$ & 5.4\,M$_{\sun}$ \\ 
Column density $N_{\rm H_2}$ & $1.6\times 10^{24}$\,cm$^{-2}$ \\ \hline
\label{tiras}
\end{tabular}
\end{table}

\section{Discussion}
\label{pdis}

As discussed in Sect.~\ref{pgps}, X-rays reduce the water abundance preferably 
in regions where the gas temperature is $T \lesssim 250$--$300$\,K 
($x({\rm H_2O}) \approx 10^{-6}$). At low densities ($n_{\rm H} \lesssim 
10^5$\,cm$^{-3}$), the fractional water abundance is less than $\approx 10^{-7}$ 
for X-ray fluxes $F_{\rm X} \gtrsim 10$\,ergs\,s$^{-1}$\,cm$^{-2}$ even for high 
gas temperatures (Fig.~\ref{ltnhx101}). For higher densities and temperatures $T 
\gtrsim 250$--$300$\,K, water has fractional abundances $x({\rm H_2O}) \gtrsim 
10^{-4}$ (nearly) independent of the X-ray flux. In the following, we discuss 
these findings in the context of protostellar envelopes and other models.

\subsection{Envelope models}

For high-mass YSOs, the observed water abundance can be successfully 
explained within the hot-core scenario in which the high temperatures in the 
inner envelope drive most of the oxygen into water and in which most of the 
water is frozen out onto grains in the outer region (e.g., Doty et al. 
\cite{dv02}; Boonman et al. \cite{bd03}). Recent interferometer studies of 
H$_2^{18}$O by van der Tak et al. (\cite{vw05}) towards the high-mass young 
stellar object (YSO) AFGL $2591$ confirm the existence of a hot ($T>100$\,K) 
compact central region with high abundant water ($x({\rm H_2O}) \approx 
10^{-4}$) and a region with lower abundances in the colder outer part. 
Although X-ray models predict the destruction of water also in high-mass 
objects (St\"auber et al. \cite{sd05}), these envelopes are likely to have 
regions with temperatures exceeding $T=250$\,K (van der Tak et al. 
\cite{ve00}). In addition, the distance from the central source to the region 
in the envelope where the gas temperature is $T=100$\,K, is much larger in 
high-mass sources compared to low-mass objects due to their high bolometric 
luminosities. The X-ray flux decreases rapidly with $r^{-2}$ and may already be 
too low to destroy H$_2$O. Water can therefore have fractional abundances of 
the order of $10^{-4}$ in envelopes around high-mass YSOs despite their 
possible X-ray emission. 

Similar jumps in the water abundances are also observed toward low-mass Class 
$0$ sources, although the derived abundances in the inner part are usually 
much lower compared to the massive objects. Ceccarelli et al. (\cite{cc00}) 
derived water abundances of $x_{\rm in}({\rm H_2O}) = 3 \times 10^{-6}$ for the 
inner part of the envelope of the Class $0$ source IRAS $16293$--$2422$ and 
$x_{\rm out}({\rm H_2O}) = 5 \times 10^{-7}$ for the outer part from detailed 
models of ISO observations. Maret et al. (\cite{mc02}) found 
$x_{\rm in}({\rm H_2O}) = 5\times 10^{-6}$ for the Class $0$ object 
NGC$1333$--IRAS$4$ and $x_{\rm out}({\rm H_2O}) = 5\times 10^{-7}$. Within 
the spherical envelope models in Sect.~\ref{pcl0}, such jumps with 
relatively low inner water abundances occur for high X-ray luminosities 
$L_{\rm X} = 10^{31}$--$10^{32}$\,ergs\,s$^{-1}$ and $t=10^5$\,yrs 
(Fig.~\ref{ih2ohx2}). X-rays are therefore a possible explanation for the low 
water abundances found in these objects.

Class I envelopes may have no or only a small region where the temperature 
exceeds $T \approx 250$\,K. The inner part of the envelope may already have 
been dispersed by outflows or accreted onto the star/disk system. Water is 
therefore easily destroyed by X-rays. According to the models in 
Sect.~\ref{prx}, the average water abundance is predicted to be at most 
$x({\rm H_2O}) \approx 10^{-6}$ in Class I sources with X-ray luminosities 
$L_{\rm X} \gtrsim 10^{27}$\,ergs\,s$^{-1}$. Note that the generic Class I 
model in Sect.~\ref{pmp} is considered as an extreme case where the gas 
temperature gets $T\approx 250$\,K close to the protostar. If the inner 
envelope has cavities of any kind, the gas temperature of the envelope will 
be lower at the inner edge of the envelope ($r_{\rm in}$), since this will be 
at a larger distance and since the dust temperature roughly falls with $r_{\rm 
in}^{-0.4}$. Water will in that case be destroyed even more readily 
(Sect.~\ref{prx}, Fig.~\ref{h2o50au}), unless there is some mechanism (e.g., 
photodesorption) that returns water ice to the gas phase.

\subsection{Disks and Outflows}

As discussed in Sect.~\ref{ptmc1}, the inner envelopes of Class $0$ and I 
sources are likely to have a much more complex geometry than the spherical 
envelopes adopted here. If protostellar cavities or holes (J{\o}rgensen et al. 
\cite{jl05}) are common, the emission from H$_2$O and other typical "hot core" 
species may be dominated by the protostellar accretion disk or outflow hot 
spots rather than by the envelope. Disks are known to be present (e.g., Brown et 
al. \cite{bc00}; Looney et al. \cite{lm00}) and can dominate over the hot core in 
terms of column density (J{\o}rgensen et al. \cite{jb05}). From the chemical 
point of view, the disk and hot core scenario are difficult to distinguish since 
both have high densities and temperatures in their inner region. According to the 
results in Sect.~\ref{pgps}, the X-ray flux of a protostar with $L_{\rm X} = 
10^{31}$\,ergs\,s$^{-1}$ and an absorbing column density of $N_{\rm H} = 
10^{24}$\,cm$^{-2}$ at the outer part of a protostellar disk ($r\approx 
400$\,AU) is $F_{\rm X} \approx 10^{-3}$\,ergs\,s$^{-1}$\,cm$^{-2}$. Water at 
temperatures below $T \approx 250$\,K is therefore destroyed from initial 
$x({\rm H_2O}) \approx 10^{-4}$ down to $x({\rm H_2O}) \approx 
10^{-7}$--$10^{-6}$. High column densities and/or temperatures are needed 
for the water to persist at $x({\rm H_2O}) \approx 10^{-4}$ in protostellar 
disks. For comparison, the chemistry disk models including X-rays by 
Aikawa \& Herbst (\cite{ah01}) have H$_2$O abundances below $\approx 10^{-8}$ 
at $r=500$\,AU from the central source. In the disk atmosphere models of 
Glassgold et al. (\cite{gn04}), water has fractional abundances $x({\rm H_2O}) 
< 10^{-5}$ at $1$\,AU from the source for vertical column densities of less 
than a few $\times 10^{23}$\,cm$^{-2}$.

Recent results presented by Chandler et al. (\cite{cb05}) for the inner 
$\approx 1$\arcsec\, of IRAS $16293$--$2422$ suggest outflow shocks to be the 
origin for the hot core species. Giannini et al. (\cite{gn01}) and Nisini 
et al. (\cite{ng02}) pointed out the importance of outflows for the water 
abundances in Class $0$ and I sources on larger scales. The results of 
Sect.~\ref{pgps} imply that the source of H$_2$O emission with $x({\rm 
H_2O}) \approx 10^{-4}$ from outflow regions with gas temperatures $100$\,K 
$\lesssim T\lesssim 250$\,K has to be located at a certain distance 
$d(L_{\rm X})$ from the X-ray source in order to survive. Consider, for 
example, $t=10^4$\,yrs, typical outflow densities ($n_{\rm H} = 
10^4$--$10^6$\,cm$^{-3}$) and an X-ray absorbing column density $N_{\rm H} = 
5\times 10^{21}$\,cm$^{-2}$. It can then be concluded from 
Figs.~\ref{ltnlx93}--\ref{ltnhx101} (Sect.~\ref{pgps}), that the H$_2$O emitting 
source would have to be at $d \gtrsim 170$\,AU for $L_{\rm X} = 
10^{28}$\,ergs\,s$^{-1}$, at $d \gtrsim 550$\,AU for $L_{\rm X} = 
10^{29}$\,ergs\,s$^{-1}$, at $d \gtrsim 1700$\,AU for $L_{\rm X} = 
10^{30}$\,ergs\,s$^{-1}$ and at $d \gtrsim 5500$\,AU for $L_{\rm X} = 
10^{31}$\,ergs\,s$^{-1}$. Larger hydrogen densities and X-ray absorbing column 
densities, however, decrease these distances. The physical quantities of the 
observed outflows by Benedettini et al. (\cite{bv02}) range between $300$\,K 
$\lesssim T \lesssim 1400$\,K and $4\times 10^4$\,cm$^{-3}$ $\lesssim n_{\rm H} 
\lesssim 2\times 10^6$\,cm$^{-3}$. It is interesting to see, that the 
derived water abundances for $T>300$\,K by Benedettini et al. are a factor 
of $\approx 10$ higher than those at $T=300$\,K. This is in good agreement 
with the results in Sect.~\ref{pgps} where the X-ray models show a jump at 
these densities but where the models without X-rays show no jump at $T=300$\,K. 
X-rays are therefore also predicted to influence the water abundances in 
outflow hot spots. It should be noted that FUV fields from the central star may 
also influence the water abundances if the photons can impact the H$_2$O 
emission region relatively unhindered, e.g. through outflow cones. 
High FUV fields tend to destroy water even at high temperatures though 
(Sect.~\ref{prfuv}; St\"auber et al. \cite{sd04}).

\section{Conclusion}
\label{pc}

The gas-phase water abundance is found to be critically dependent on time, gas 
temperature, hydrogen density and X-ray flux (Sect.~\ref{pgps}). Three distinct 
regimes are identified: In the first regime ($T \lesssim 100$\,K), the water 
abundance is $x({\rm H_2O}) \approx 10^{-7}$--$10^{-6}$ and can be somewhat 
enhanced or destroyed by X-rays in comparison to models without X-rays. In the 
second regime ($250$\,K $\lesssim T \lesssim 100$\,K), water is released from 
grains with $x({\rm H_2O}) \approx 10^{-4}$ but quickly reduced by X-rays to 
fractional abundances $x({\rm H_2O}) \approx 10^{-6}$. The 
third regime ($T \gtrsim 250$\,K) is characterized by high water abundances 
($x({\rm H_2O}) \gtrsim 10^{-4}$) due to the efficient reaction of OH with 
molecular hydrogen. At low densities ($n_{\rm H} \lesssim 10^{5}$\,cm$^{-3}$) 
water is destroyed even at high temperatures for X-ray fluxes $F_{\rm X} 
\gtrsim 10$\,ergs\,s$^{-1}$\,cm$^{-2}$. In general, higher gas temperatures and 
higher hydrogen densities allow higher X-ray fluxes for H$_2$O to survive. 
Water is mainly destroyed in reactions with the X-ray enhanced species HCO$^+$ 
and H$_3^+$. It is also destroyed by internally created FUV photons.  

The envelopes of both Class I (Sect.~\ref{pmp}) and Class $0$ 
(Sect.~\ref{pcl0}) objects were modeled under the influence of central X-ray 
and FUV emission. The results for the Class I envelope show that an initial 
warm gas-phase water abundance of $x({\rm H_2O}) \approx 10^{-4}$ is reduced to 
$x({\rm H_2O}) \approx 10^{-6}$ within $\approx 5000$\,yrs for X-ray 
luminosities $L_{\rm X} \gtrsim 10^{28}$\,ergs\,s$^{-1}$ and within 
$\approx 5\times 10^4$\,yrs for X-ray luminosities $L_{\rm X} \gtrsim 
10^{27}$\,ergs\,s$^{-1}$, consistent with reported upper limits for H$_2$O 
towards Class I sources. Water is also destroyed by X-rays in the Class $0$ 
object, but higher X-ray fluxes are needed due to the higher densities in the 
inner region. The influence of a central FUV field is negligible in our models, 
unless the FUV photons can escape to larger distances (Sect.~\ref{prfuv}).

The current protostellar models are limited to spherical symmetry, which 
is obviously an approximation to the true geometry of the inner few hundred 
AU where cavities and flattened disks may be present. Nevertheless, the high 
densities and temperatures in the current models are representative of such 
regions. In particular, X-rays are predicted to regulate the H$_2$O abundances 
wherever the gas temperature is $T \lesssim 250$--$300$\,K. This will not only 
be the case for H$_2$O originating from envelopes or disks in high-mass 
star-forming regions, but also for H$_2$O produced in outflow hot spots around 
low-mass objects. The presence of cavities or outflow cones will have the 
result that X-rays and/or FUV photons penetrate to larger radii in certain 
directions.

Future instruments such as Herschel Space Observatory will allow detailed 
studies of the water abundances in both high and low-mass YSOs. With its high 
resolution instrument HIFI (de Graauw \& Helmich \cite{dh01}), it will be 
possible to distinguish shock heated H$_2$O from H$_2$O in hot cores or disks 
from their line profiles. Observations of the optically thin isotope H$_2^{18}$O 
will help to sample the envelope, whereas optically thick lines may give 
information about outflow properties. PACS will delineate the water abundances 
on $9$\arcsec\, scale including maps of the outflows. ALMA, on the other hand, 
will be able to resolve the innermost region ($<0.1$\arcsec) of YSO envelopes 
and/or disks in order to study the chemistry and therefore its physical properties 
in molecules other than H$_2$O. Limited maps of H$_2^{18}$O on arcsec scale may 
be possible though with ALMA under exceptional conditions in the $203$\,GHz line. 
High-$J$ lines of CO will reveal the gas temperature in these regions whereas the 
observation of X-ray and FUV tracers (St\"auber et al. \cite{sd04}, \cite{sd05}) 
may help to clarify the high-energy properties of young stellar objects. Together, 
these new facilities and models will reveal much about the physical and 
chemical structure of the inner regions of young stellar objects, which are 
currently poorly understood.


\begin{acknowledgements}

This work was partially supported under grants from The Research Corporation 
(SDD). The research of JKJ was supported by NASA Origins Grant NAG5-13050. 
Astrochemistry in Leiden is supported by the Netherlands Research School 
for Astronomy (NOVA) and by a Spinoza grant from the Netherlands Organization 
for Scientific Research (NWO).

\end{acknowledgements}



\end{document}